\documentclass[prb, showpacs, superscriptaddress]{revtex4}

\usepackage{graphicx}%

\usepackage{amssymb}
\usepackage{amsmath}
\usepackage{mathtools}

\begin{document}

\title{Close packed structure dynamics with finite range interaction: computational mechanics with individual layer interaction}

\author{Edwin \surname{Rodriguez-Horta}}
\author{Ernesto \surname{Estevez-Rams}}
\affiliation{Facultad de F\'isica-IMRE, Universidad de la Habana, San Lazaro y L. CP 10400. C. Habana. Cuba}
\email{estevez@fisica.uh.cu}
\author{Raimundo \surname{Lora-Serrano}}
\affiliation{Universidade Federal de Uberlandia, AV. Joao Naves de Avila, 2121- Campus Santa Monica, CEP 38408-144, Minas Gerais, Brazil}
\author{Reinhard \surname{Neder}}
\affiliation{Kristallographie und Strukturphysik. Universit\"at Erlangen. Germany}

\date{\today}

\begin{abstract}
This is the second contribution in a series of papers dealing with dynamical models in equilibrium theories of polytypism. A Hamiltonian introduced by \cite{ahmad00} avoids the unphysical assignment of interaction terms to fictitious entities given by spins in the H\"agg coding of the stacking arrangement. In this paper an analysis of polytype generation and disorder in close packed structure is made for such Hamiltonian. Results are compared with a previous analysis using the Ising model. Computational mechanics is the framework under which the analysis is performed. The competing effects of disorder and structure, as given by entropy density and excess entropy, respectively, is discussed. It is argued that the Ahmad and Khan model is simpler and predicts a larger set of polytypes than previous treatments. 
\end{abstract}


\maketitle

\section{Introduction}

In the study of the origins of polytypism, equilibrium theories relying on the Ising models have played a relevant role ( See \cite{verma66} and references therein, also  \cite{cheng88}, \cite{kabra88}, \cite{engels90}, \cite{shaw90}). To code the stacking ordering in such a way that is suitable for the Ising model, the usual procedure has been to assign a spin (up or down) to a pair of consecutive layers in the stack. This assignment is done  according to their arrangement: one spin, say up, for layers stacked in clockwise or cyclic manner (related by a $60^o$ rotation around a common axis), and the other spin, say down, for the opposite ($-60^o$ rotation around a common axis). This is nothing else than the H\"agg or the Nabarro-Frank coding of layer arrangement  \cite{verma66}. In a previous article \cite{edwin16} (from now on RENL16) we analyzed, through the light of computational mechanics, such Ising models which will be referred to as (such models will be referred to as Ising model between layer pairs or IMLP), enhancing the usual treatment based on statistical physics. The analysis allowed to study, using a second neighbor interaction, not only the topology of the phase diagram, but also the occurrence of disorder and longer length polytypes in a region at the boundary between phases. Quantitative assessment of the balance between disorder and pattern was achieved through the use of magnitudes such as entropy density and excess entropy.

A troubling aspect of the IMLP, as used in RENL16 or those before, is the rather artificial assignment of an interaction term between fictitious entities such as pair of layers. This has been mostly overlooked in the literature. When it has been done, the approach has been to introduce interaction terms between an even number of spins to go beyond pairwise interactions and partially account for the problem \cite{cheng88,engels90}. An exception has been the contribution of \cite{ahmad00}, where the usual Ising Hamiltonian is dropped all together and, instead, a model is built with the interaction taken between pair of individual, physically real, layers. In this contribution we set to further explore the Ahmad and Khan model by using computational mechanics. 

Computational mechanics draws from the realm of information theory to study quantitatively the emergence of pattern and complexity in dynamical systems. It has been already used in the study of polytypism \cite{varn02,varn04,varn06,varn13,varn13a,riechers14}. The reader is kindly advised to read our previous paper RENL16 for a more thorough discussion of the mathematics involved in the use of computational mechanics in the current subject, as in this paper we will use the same formalism now applied to the Ahmad and Khan Hamiltonian (AKH).    

The paper is organized as follows: in Section \ref{sec:math} the AKH model is presented and the difference with the conventional Ising model discussed. In Section \ref{sec:NNN} the mathematics of the model within computational mechanics is presented, this is followed in Section \ref{sec:discussion} by a discussion of the results and finally, the conclusions. 

\section{Interaction model between individual layers.\label{sec:math}}

Close packed stacking arrangements are usually described by three letters $A$, $B$ and $C$, corresponding to three displacement in the plane of the layer with respect to a common fixed origin \cite{patterson}. Close packed means that consecutive layers can not be in the same lateral position, that is, the same letter can not be found consecutive in the description of the stacking order. If the layer $A$ is taken to represent zero displacement, then, using $\vec{a}$ and $\vec{b}$ as the basis vectors for the hexagonal plane lattice, layer $B$ could corresponds to a   displacement of $\vec{q}=1/3 \vec{a}+2/3 \vec{b}$, and $C$ to a displacement of  $-\vec{q}=2/3 \vec{a}+1/3 \vec{b}$. With this description when going from one layer to the next one, two possibilities arises: in one case the arriving layer is displaced a vector $\vec{q}$ with respect to the starting layer; in the other case the displacement is of $-\vec{q}$. To the first possibility a spin up is assigned ($1$), while in the second case the opposite is taken $(-1)$. This coding completely specifies the stacking arrangement if the initial layer is specified. The coding is known as the H\"agg code when the sign of the spin ($+$ or $-$) is used, and Nabarro-Frank code when $\bigtriangleup$ and $\bigtriangledown$ symbols are used. 

In RENL16, the Ising model is written for the spin coding using the usual Hamiltonian
\begin{equation}
  \displaystyle H=-\sum\limits_{k}^{n}J_{k}\sum\limits_{j} s_{j} s_{j+k},\label{eq:isinghamiltonian}
\end{equation}
where $J_k$ is the interaction parameter for range $k$, and $s_i$ is the spin assigned to a pair of layers at site $i$. The interaction range is $n$.

In spite of its beauty, the described Ising Hamiltonian raises some serious concerns. First of all, there is no physical basis for the type of interactions it proposes between two layer displacement. The interaction between the double layer $CA$ and $BC$ is positive, while between $CA$ and $AC$ is negative without any physical reason to justify it. The second reason is that the orientation between one layer and its neighbor, determined by the assigned spin, contains, for sufficiently separated layers, no information about the interacting entities. Trying to reinterpret the spin interaction as some kind of interaction between individual layers turns to be inconsistent. For example, consider the sequence $CACABA$, which corresponds to the H\"agg coding $+-++--$. The interaction between the first layer $C$ and the fourth layer $A$ contributes with a positive term, whereas the third layer $C$ and the six layer $A$ results in a negative term. For the same combination of layers, at the same distance, opposite interaction terms are found depending on the intermediate layers between them.  Yet, pairwise interaction should depend on the nature of the layers themselves and the displacement between them and therefore, in the above example should be the same for both cases. 

In order to surmount this difficulty  \cite{shaw90} introduced into the Ising Hamiltonian collective terms involving an even number of layers. The number of involved layers in the additional interaction terms must be even to obey the space reversion symmetry. The resulting Hamiltonian, although mathematically sounder, is cumbersome and even more physically bizarre to explain in terms of real interaction forces and the corresponding energies. 

Another point of view was taken by \cite{ahmad00}. By getting rid of the H\"agg coding, they wrote a Hamiltonian of the type
\begin{equation}
\displaystyle  H=\sum\limits_{i=1}^{m}\left(A_i\sum\limits_{k=1}^{N}S_{ki}\right),\label{eq:ahmadhamiltonian}
\end{equation}
where the $A_i$ are now the interaction parameters and  $S_{ki}$ takes value +1 if layer $k$ and $k+i$ are in the same orientation ($A-A$, $B-B$ or $C-C$), or $-1$ otherwise. Now $m$ is the interaction range. In this model, no spins are assigned to pair of consecutive layers depending on their displacement. For the above stacking arrangement $CACABA$, interaction between the first and fourth layer has the same contribution as the interaction between the third and six layer. The AKH given by (\ref{eq:ahmadhamiltonian}) is truly about interaction between individual layers. 

The reader should notice that an IMLP of order $n$ actually represents an interaction that goes to layer separated by a distance up to $n+1$, so this has to be taken into account when comparing with the AKH. Therefore the second order IMLP is compared with the third order AKH. Also, in the IMLP, the close packed constrain is already coded within the spin assignment. In the AKH model no. As two consecutive layers can not be in the same orientation, the contribution of the $A_1$ term is constant and equal for any arrangement of the same length, therefore $A_1$ can be taken as zero. 

One may ask if both models are topologically equivalent and if there exist a one-to-one and therefore invertible relation between them. This is not the case. In order to understand this consider the second order Ising term $J_2$, which codes interaction between pairs of layers $s_i$ and pair of layers $s_{i+2}$ in a sequence of the type $X_1X_2X_3X_4$, where $X_i$ stands for some letter $A$, or $B$ or $C$. It can be seen that, as long as $S_{k3}$ equals $-1$, both contributions to the Hamiltonian will be the same if $J_2=A_3$, but if $S_{k3}=1$ then this is no longer the case. 

Consider the sequences of $N$ layers corresponding to the perfect hexagonal close packed (HCP) polytype ($ABABAB\cdots$), the perfect face centered (FCC) polytype ($ABCABC\cdots$) and the perfect double hexagonal close packed (DHCP) polytype ($ABCBABCB\cdots$). If the energy for each phase using the IMLP and the AKH are calculated, the equivalence of the interaction parameters would be
\begin{equation}
 \begin{array}{l}
  J_1=-A_2 \; \text{for all three stacking arrangements}\\
  J_2=\left\{ \begin{array}{ll}
  -A_3 & \text{for HCP} \\
  A_3 & \text{for FCC and DHCP}
  \end{array}\right.\\
 \end{array}
\end{equation}
This simple example shows that there is no invertible correspondence between the parameters of both models and therefore no conformal mapping between the corresponding phase diagrams.

\section{Stacking sequence as a pairwise layer interaction model in one dimension\label{sec:NNN}}

In what follows the mathematical framework developed in reference RENL16 will be followed.

For the AKH model up to third order interaction, the total energy of the system is given by
\begin{equation}
H=A_2\sum\limits_{k=1}^{N}S_{k2}+A_3\sum\limits_{k=1}^{N}S_{k3}.\label{eq:ahmadhamiltonian3}
\end{equation}
The string of letters describing the arrangement can be split in non overlapping blocks of length three, making the problem a Markovian system. The ordered set of $\eta$ of possible blocks are
\[\begin{array}{rl}\eta=&\{ABA,ABC,ACA,ACB,BAB,BAC,\\\\&BCA,BCB,CAB,CAC,CBA,CBC\}.\end{array}\]
To the letters $A$, $B$ and $C$ the numbers $0$, $1$ and $2$ are assigned, respectively. With that notation, the interaction variable $S_{ki}$ takes the form 
\begin{equation}
 S_{ki}=|l_k-l_{k+i}|^2-3|l_k-l_{k+i}|+1,\label{eq:ski}
\end{equation}
where $l_k$ is the $k$ layer in the sequence. From the Hamiltonian (\ref{eq:ahmadhamiltonian3}) and equation (\ref{eq:ski}) the transfer matrix can be built. In doing so, care must be taken to entries corresponding to adjacent blocks that violates the close packed condition (e.g $ABA$ followed by $ABC$). These entries must be set to a very high energy value so the resulting probability of occurrence is effectively zero. 

The central idea of Computational Mechanics is to build the optimal computational machine able to reproduce the statistical behavior of the system, seen as a dynamical process. In our case, the class of machines sought are finite state machines (FSM) that outputs, in discrete steps, equal length blocks of symbols. For sufficiently large number of steps, the output string must be statistically representative of the stacking arrangement. At any given step, the machine is in a state of the FSM and will jump to another state as it outputs a block of symbols.  The probability of jumping from one state to another can be described by a stochastic matrix that depends on the system parameters $A_2$ and $A_3$, and the Boltzmann factor $\beta=1/T$. In RENL16, the derivation of the stochastic matrix from the transfer matrix formalism is described.

Once the stochastic matrix is found, the minimum finite state machine, within the class of machines explored, that optimally predicts the system behavior can be constructed. In such machine, termed $\epsilon$-machine, the states are named causal states. The $\epsilon$-machine description allows to calculate several parameters related to disorder, such as entropy density $h_\mu$, which measures the amount of random behavior of the system. It also allows to measure pattern or structure, given by the excess entropy $E$ or mutual information between any two substring partition of the whole stacking arrangement taken of infinite length. The amount of resources (memory) the system needs to statistically reproduce its behavior can also be calculated. It will be given by the statistical complexity $C_\mu$ or the Shannon entropy over the causal states. The reader is refered to RENL16 and reference therein for further discussion of these parameters. 

To start, consider the ground state corresponding to $\beta\rightarrow \infty$. The interaction range implies at most twelve causal states in the corresponding FSM describing the dynamics (see supplementary material). This generic FSM will have the largest possible connectivity (MC-FSM). Yet, at $T=0$ some states are unreachable. Other states have identical type of rows in their stochastic matrix. This rows represent the same causal state and therefore collapse to a single state in the $\epsilon$-machine representation. 

For $A_3/A_2<1$ and $A_2>0$ the transition matrix between states in the FSM description is given by
\begin{equation}
 \displaystyle P=\left ( \begin{array}{cccccccccccccc}
			  0 & 0 & s & 0 & s & 0& s & s & 0 & s & 0& s & 0\\
			  0 & 0 & 0 & 0 & 0 & 0& 0 & 0 & 0 & 0& 0& 0& 0\\
			  0 & 0 & 1 & 0 & 0 & 0& 0 & 0 & 0 & 0& 0& 0& 0\\
			  0 & 0 & 0 & 0 & 0 & 0& 0 & 0 & 0 & 0& 0& 0& 0\\
			  0 & 0 & 0 & 0 & 1 & 0& 0 & 0 & 0 & 0& 0& 0& 0\\
			  0 & 0 & 0 & 0 & 0 & 0& 0 & 0 & 0 & 0& 0& 0& 0\\
			  0 & 0 & 0 & 0 & 0 & 0& 1 & 0 & 0 & 0& 0& 0& 0\\
			  0 & 0 & 0 & 0 & 0 & 0& 0 & 1 & 0 & 0& 0& 0& 0\\
			  0 & 0 & 0 & 0 & 0 & 0& 0 & 0 & 0 & 0& 0& 0& 0\\ 
			  0 & 0 & 0 & 0 & 0 & 0& 0 & 0 & 0 & 1& 0& 0& 0\\ 
			  0 & 0 & 0 & 0 & 0 & 0& 0 & 0 & 0 & 0& 0& 0& 0\\
			  0 & 0 & 0 & 0 & 0 & 0& 0 & 0 & 0 & 0& 0& 1& 0\\
			  0 & 0 & 0 & 0 & 0 & 0& 0 & 0 & 0 & 0& 0& 0& 0\\
			  \end{array}
\right ),
\end{equation}
where $s=1/6$. Each entry $p_{ij}$ in the transition  matrix is the probability of making a transition from state $i$ to state $j$. The states are given by the ordered set $\eta$, where a starting state $s$ is prepend to the set.

\begin{figure}
	\centering
	\includegraphics[scale=0.8]{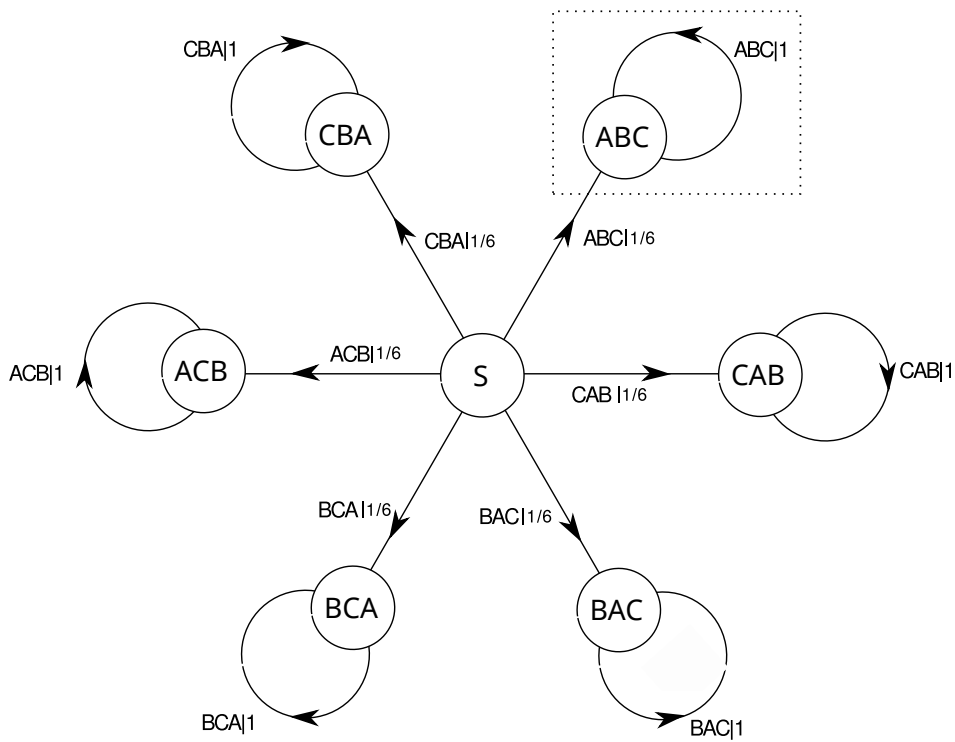}
	\caption{The Finite State Machine (FSM) describing the FCC polytype at $\beta\rightarrow\infty$. The transient initial state is connected with six recurrent states with uniform probability of $1/6$. The recurrent states are isolated, once the system leaves the starting state it will remain in the recurrent state it has chosen. The $\epsilon$-machines is therefore just one if the recurrent states and one instance is shown within a dashed box.}
	\label{fsm_nnn_3c}
\end{figure}

The analysis of the above matrix shows that only six states are recurrent (stationary probability different from zero): $ABC$, $ACB$, $BAC$, $BCA$, $CAB$, $CBA$. From the starting state $S$, once a recurrent state is reached, the system stays there (Fig. \ref{fsm_nnn_3c}). Each state represents an equivalent FCC ($3C$) stacking ordering. All the states are therefore equivalent and redundant. The stationary $\epsilon$-machine consist of a single causal state (FCC-FSM), which can represent any of the six $3C$ states. The system is one where the output block is always the same with probability one. Statistical complexity is $C_{\mu}=0$ bits and entropy density is  $h_{\mu}=0$ bits/site. 

\begin{figure}
	\centering
		\includegraphics*[scale=0.6]{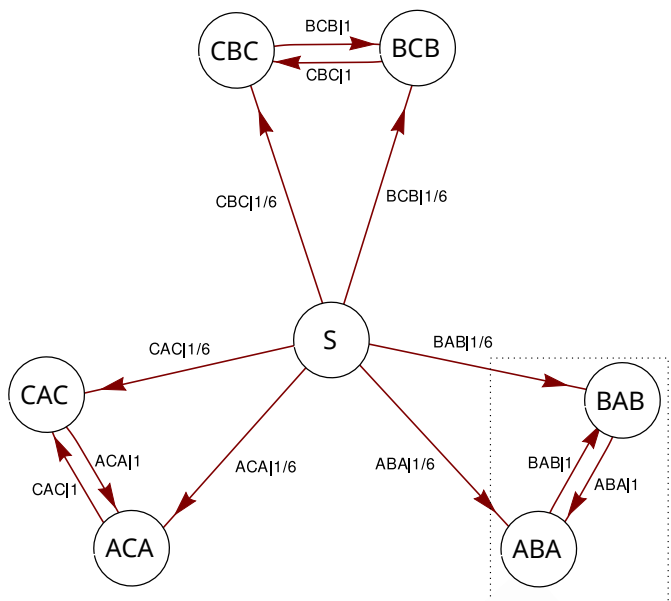}
	\caption{The FSM describing the HCP polytype at $\beta\rightarrow\infty$. The transient initial state is connected with six recurrent states with uniform probability of $1/6$. The six recurrent states are connected by pairs, once the systems has left the starting state it will alternate between two connected states. The $\epsilon$-machines is therefore just one pair of connected recurrent states and one instance is shown within a dashed box..}
	\label{fsm_nnn_2h}
\end{figure}

Consider $A_3/A_2<1$ and $A_2<0$, the transition matrix in this case follows 
\begin{equation}
\displaystyle P=\left ( \begin{array}{cccccccccccccc}
			0 & s & 0 & s & 0 & s & 0 & 0 & s & 0 & s & 0& s\\
			0 & 0 & 0 & 0 & 0 & 1& 0 & 0 & 0 & 0& 0& 0& 0\\
			0 & 0 & 0 & 0 & 0 & 0& 0 & 0 & 0 & 0& 0& 0& 0\\
			0 & 0 & 0 & 0 & 0 & 0& 0 & 0 & 0 & 0& 1& 0& 0\\
			0 & 0 & 0 & 0 & 0 & 0& 0 & 0 & 0 & 0& 0& 0& 0\\
			0 & 1 & 0 & 0 & 0 & 0& 0 & 0 & 0 & 0& 0& 0& 0\\
			0 & 0 & 0 & 0 & 0 & 0& 0 & 0 & 0 & 0& 0& 0& 0\\
			0 & 0 & 0 & 0 & 0 & 0& 0 & 0 & 0 & 0& 0& 0& 0\\
			0 & 0 & 0 & 0 & 0 & 0& 0 & 0 & 0 & 0& 0& 0& 1\\ 
			0 & 0 & 0 & 0 & 0 & 0& 0 & 0 & 0 & 0& 0& 0& 0\\ 
			0 & 0 & 0 & 1 & 0 & 0& 0 & 0 & 0 & 0& 0& 0& 0\\
			0 & 0 & 0 & 0 & 0 & 0& 0 & 0 & 0 & 0& 0& 0& 0\\
			0 & 0 & 0 & 0 & 0 & 0& 0 & 0 & 1 & 0& 0& 0& 0\\
			\end{array}
\right ),
\end{equation}
again $s=1/6$. Possible configurations can only be generated by the pair of causal states: ABA-BAB, ACA-CAC, BCB-CBC. The six involved states are referred to as the HCP or 2H states. Each pair results in a $2H$ sequence. Once, from the starting state, one of the pair of connected causal state is reached, the system stays there (Fig. \ref{fsm_nnn_2h}). The corresponding stationary $\epsilon$-machine (HCP-FSM) consist on one pair of connected causal states. Statistical complexity is $C_{\mu}=1$ bits and entropy density is $h_{\mu}=0$ bits/site. 

\begin{figure}
	\centering
	\includegraphics*[scale=0.5]{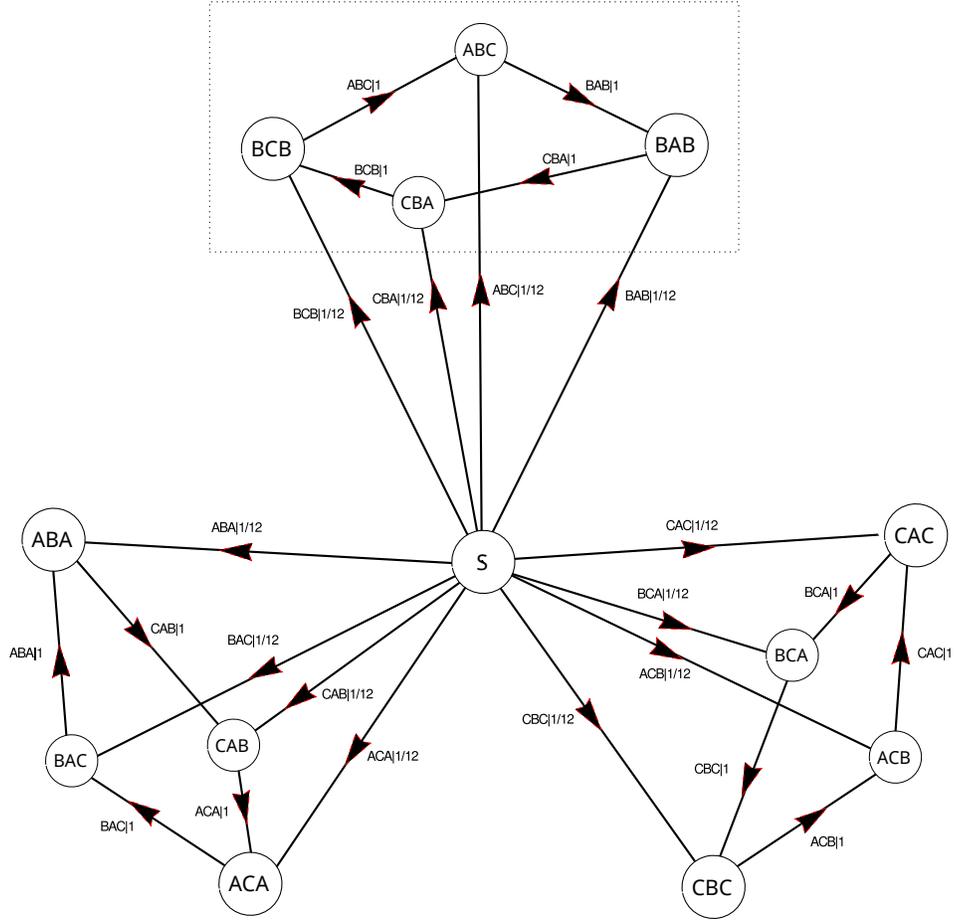}
	\caption{The FSM describing the DHCP polytype at $\beta\rightarrow\infty$. The initial states is connected uniformly with twelve recurrent states which are grouped in three sets of four connected states each. The groups are isolated between them. The system leaves the starting states to one state in one of the groups and then it will walk with absolute certainty in a given fixed order through the states in the group, resulting in a cyclic transition between those states. The $\epsilon$-machine will be described by any one of the groups and one instance is shown within a dashed box..}
	\label{fsm_nnn_4h}
\end{figure}

Finally, another region is defined by $A_3/A_2>\frac{1}{2}$ and $A_2>0$, where the transition matrix for the FSM has the form:
\begin{equation}
\displaystyle P=\left ( \begin{array}{cccccccccccccc}
			0 & s & s & s & s & s & s & s & s & s & s & s & s\\
			0 & 0 & 0 & 0 & 0 & 0& 0 & 0 & 0 & 1& 0& 0& 0\\
			0 & 0 & 0 & 0 & 0 & 1& 0 & 0 & 0 & 0& 0& 0& 0\\
			0 & 0 & 0 & 0 & 0 & 0& 1 & 0 & 0 & 0& 0& 0& 0\\
			0 & 0 & 0 & 0 & 0 & 0& 0 & 0 & 0 & 0& 1& 0& 0\\
			0 & 0 & 0 & 0 & 0 & 0& 0 & 0 & 0 & 0& 0& 1& 0\\
			0 & 1 & 0 & 0 & 0 & 0& 0 & 0 & 0 & 0& 0& 0& 0\\
			0 & 0 & 0 & 0 & 0 & 0& 0 & 0 & 0 & 0& 0& 0& 1\\
			0 & 0 & 1 & 0 & 0 & 0& 0 & 0 & 0 & 0& 0& 0& 0\\ 
			0 & 0 & 0 & 1 & 0 & 0& 0 & 0 & 0 & 0& 0& 0& 0\\ 
			0 & 0 & 0 & 0 & 0 & 0& 0 & 1 & 0 & 0& 0& 0& 0\\
			0 & 0 & 0 & 0 & 0 & 0& 0 & 0 & 1 & 0& 0& 0& 0\\
			0 & 0 & 0 & 0 & 1 & 0& 0 & 0 & 0 & 0& 0& 0& 0\\
			\end{array}
\right ),
\end{equation}
now $s=1/12$. This matrix describes three loops of four states: $ABC-BAB-CBA-BCB$; $ACB-CAC-BCA-CBC$; $ABA-CAB-ACA-BAC$ (Fig. \ref{fsm_nnn_4h}). Each loop reproduces the DHCP configuration and, being equivalent, any of them will describe the $\epsilon$-machine for such phase. Statistical complexity $C_{\mu}=2$ bits and the entropy density $h_{\mu}=0$ bits/site.

\begin{figure}
	\centering
	\includegraphics*[scale=0.7]{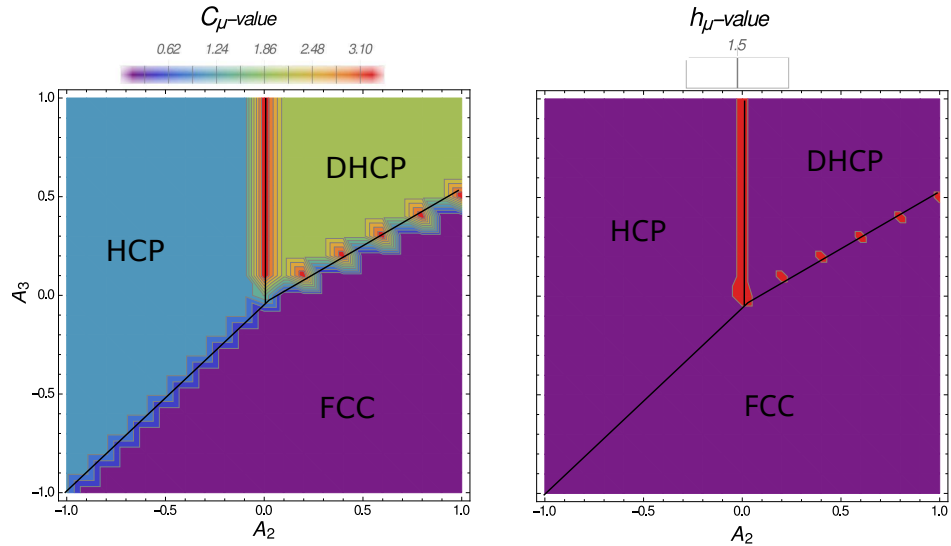}
	\caption{Statistical complexity $C_\mu$ as function of the interaction parameters $A_2\times A_3$ (left); the entropy density $h_\mu$ as a function of  $A_2\times A_3$ (right). In both plots temperature was taken as zero ($\beta \rightarrow \infty$).}
	\label{chbetainfb0}
\end{figure}

The plot of the statistical complexity versus the interaction parameters $A_2\times A_3$ reproduces the phase diagram at $\beta\rightarrow \infty$ (Fig. \ref{chbetainfb0} left). Forked shaped boundary lines can be recognized where, according to the entropy density diagram of figure \ref{chbetainfb0}-right, is the only region where disorder can be found.

Lines $A_2=0$, $A_3>0$ and $A_3=A_2/2>0$  represent boundaries between the DHCP phase and the HCP and FCC phase, respectively. For both boundaries the FSM is described by twelve causal states connected between them. Disorder is high ($h_\mu\approx2.1$ bits/site) as well as statistical complexity $C_\mu$ ($\approx3.4$ bits). The $\epsilon$-machine describing the FCC-DHCP border can be explained as the union of the FSMs describing phases FCC and DHCP. In other words, there is a combination of the self referenced states of FCC and the set of four loops found in the FSM of the DHCP. A similar analysis can be done for the HCP-DHCP. The connectivity between the FSM describing the phases outside the boundaries is responsible, at the boundary, of the emergence of new polytypes.  Following \cite{estevez08} all possible polytypes up to a length of 12 layers were generated, and their probability of occurrence calculated using the $\epsilon$-machine description. Tables I and II show all polytypes with occurrence probability above  zero. All polytypes considered were taken as result of close loops in the FSM description. Those are polytypes that can be truly considered as intrinsic to the system energetics. Other layer sequences present are result of  boundaries configurations and can not be considered as truly emerging polytypes. Probability has been normalized to the sequence length to make comparison meaningful. It is clear that the boundary lines $A_2=0$, $A_3>0$ and $A_3=A_2/2>0$ are multiphase regions, where degenerate phases coexist disorderly and long length polytypes can be found.

The line $A_2=A_3<0$ divides the HCP and FCC phases. The FSM describing the stacking arrangements in such line is also a combination of the FSM of both phases yet, in this case, no transitions are possible between them. Once the system chooses one of the configurations, it stays there, and disorder is not possible ($h_\mu=0$ bits/site). If one drop the sequential approach of the FSM and considers stacking arrangement that starts growing randomly at several ``sites'' simultaneously, the analysis could point to the occurrence of a mechanical mixture of HCP and FCC stacking that eventually collides in frozen boundaries. But at $\beta \rightarrow \infty$, the antiphase frontiers have energy larger than any of the  two stable phases and at equilibrium are forbidden. The system will choose between perfect HCP or perfect FCC and not a mixture of states. The FCC-HCP border is completely ordered ($h_\mu=0$ bits/site) for a single realization of the system dynamics.

\subsection{Stacking arrangement for temperature above zero ($\beta < \infty$)}

As soon as temperature increases above zero, the system looses order, which in turn implies a larger number of causal states involved on the description of the stacking arrangement. Causal states that were disconnected at $\beta \rightarrow \infty$, will start having, as temperature increases, transition probabilities different from zero. As a result, the MC-FSM will be the $\epsilon$-machine description, $C_\mu$ increases in all regions as well as the entropy density.

\begin{figure}
	\centering
	\includegraphics*[scale=0.65]{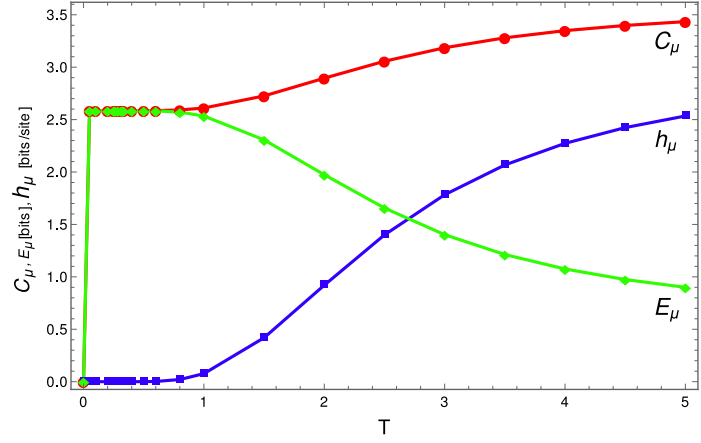}
	\caption{Statistical complexity $C_\mu$, excess entropy $E_\mu$ and entropy density  $h_\mu$ as a function of temperature $T=1/\beta$. FCC region with $A_2=1$ and $A_3=-1$.}
	\label{vsTFCC}
\end{figure}

For temperature just above zero, in the FCC region, the equivalent causal states at $T=0$ (Fig. \ref{fsm_nnn_3c}) reproducing the stacking arrangement ($ABC$, $ACB$, $BAC$, $BCA$, $CAB$, $CBA$) are now weakly linked between each other. The system will stay in one of this casual states for long periods and only, eventually, a transition will be made to another FCC state. When that happens an antiphase boundary can emerge. The occurrence of any of the FCC causal states have the same probability. In the thermodynamic limit for the infinite string ($N\rightarrow\infty$) the stacking arrangement will be made of long blocks of perfectly ordered FCC phase, that will meet at boundaries made of HCP configuration of length two. In Figure \ref{vsTFCC} the behavior of the entropic parameters with temperature is presented for the FCC region.  For small values of $T\neq0$ ($\beta \gg 1$) the stationary distribution over the six FCC causal states is uniform and the value of $C_\mu$ is $2.58$ bits. The jump in $C_\mu$ signals this change in symmetry with temperature. From there on, with increasing temperature, the probability of the 2H states stays different from zero. As temperature further increases, the stationary probability over the twelve states gets closer to uniformity as a result, statistical complexity smoothly increases up to $C_\mu=3.58$ bits where the probability over the twelve causal states becomes uniform. 

\begin{figure}
	\centering
	\includegraphics*[scale=1.0]{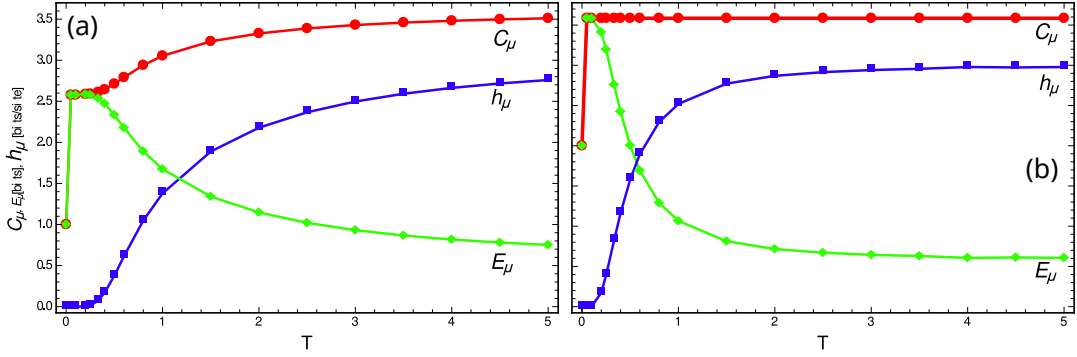}
	\caption{Statistical complexity $C_\mu$, excess entropy $E_\mu$ and entropy density  $h_\mu$ as a function of temperature $T=1/\beta$. a) HCP region with $A_2=-1$ and $A_3=1$. b) DHCP region with $A_2=1$ and $A_3=1$.}
	\label{vsT}
\end{figure}

For the HCP phase (Fig.  \ref{vsT}a), as soon as $T\neq0$ the system has a jump of the statistical complexity to $C_\mu\approx2.58$ bits due to the coupling between the six HCP state, each state with equal stationary probability. Starting at a given value of temperature (dependent on the parameters $A_2$ and $A_3$), the system entropy rate start monotonically to increase, dragging the increase of the statistical complexity and the decrease of the excess entropy. The decrease of $E$ is due to the loss of structure in the system.  For the DHCP (Fig. \ref{vsT}b) a similar behavior is observed. But now, as soon as $T\neq 0$, all twelve states are coupled and disorder is larger than in the HCP phase at the same temperature. The FCC phase is the most stable one, as can be seen by comparing the corresponding value of the entropy density for the same temperature for each phase. At $T\rightarrow \infty$ ($\beta \rightarrow 0$) the system reaches the most disordered state, but different from the model of interacting pair of layers, even at that temperature some structure remains as a result of the close packed condition. In the AKH close packed constrain must be included in the Hamiltonian and this shows up even at very high temperature.

\begin{figure}
	\centering
	\includegraphics*[scale=0.9]{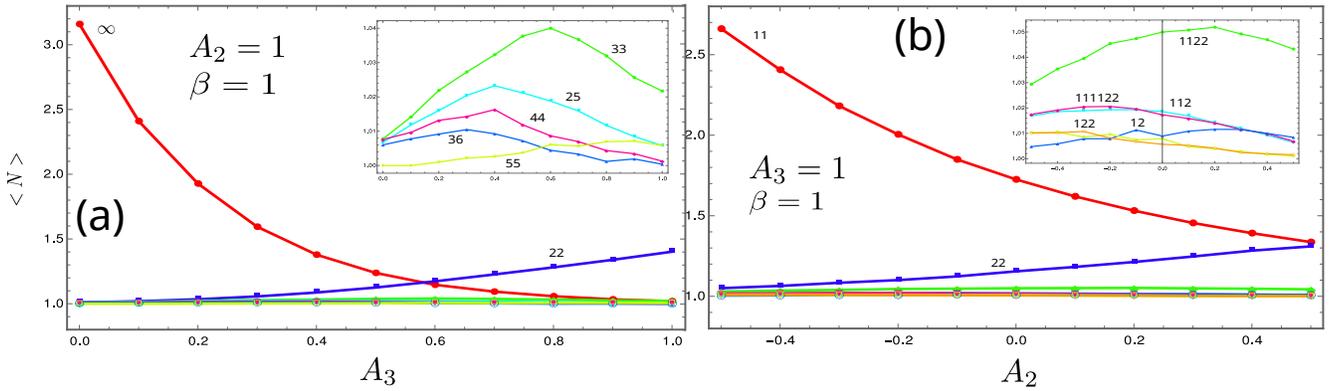}
	\caption{Average length of the polytypes in units of the polytype length as a function of (a) $A_3$ starting at the FCC phase an ending at the DHCP phase; (b) $A_2$ starting at the HCP phase an ending at the DHCP phase. Insets are the same plot at different scale showing the behavior of the frontier phases. }
	\label{NvsT}
\end{figure}

Increasing temperature also has the effect of broadening the region of existence of the phase boundary polytypes, as shown in Figure \ref{NvsT}. In the left, the average length is shown, starting well in the FCC region and moving towards the DHCP region. It can be seen that the average unperturbed sequences of FCC decrease their length, while the DHCP length increases. More interesting is that the boundary phases now exist in a band around the different phase boundaries. The same appearance of bands of existence for longer period polytypes can be seen when going from the HCP region, in the phase diagram, to the DHCP region (Figure \ref{NvsT}b). In this case all boundary phases corresponds to antiphase sequences that happens isolated at the frontier between two stable sequence as discussed before.

\begin{figure}
	\centering
	\includegraphics*[scale=1.1]{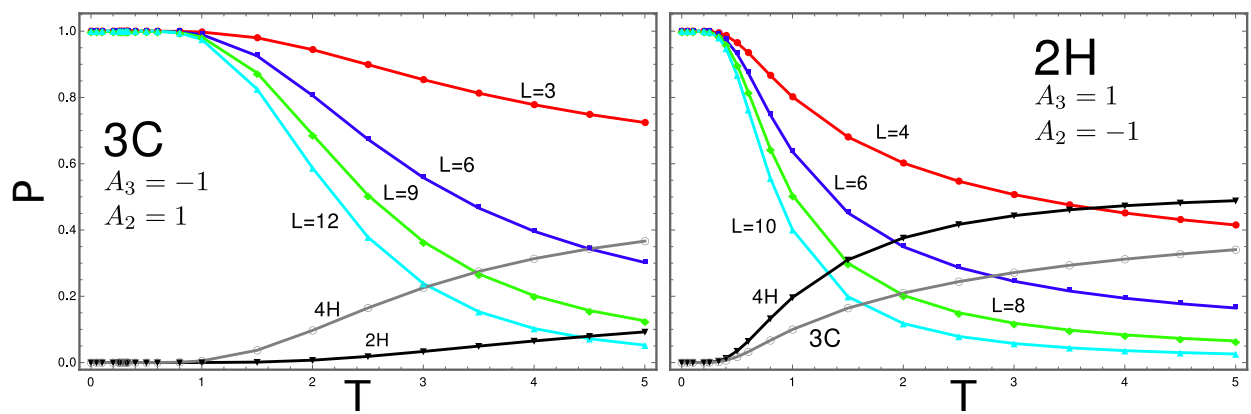}
	\caption{Probabilities of different length L of phases (3C) FCC and (2H) HCP as a function of temperature. The probability of the DHCP (L=4), FCC(L=3) and HCP (L=4) phases are shown in each plot for comparison.}
	\label{ProbvsT}
\end{figure}

Finally, the probability of different sequences were studied as a function of temperature for a point inside the FCC region (Fig. \ref{ProbvsT}-left) and the HCP region (Fig. \ref{ProbvsT}-right). As expected, the probability of all stable sequences fall with temperature, but at different speed for different sequence length of the same phase. It can be noted that the FCC phase is more robust to disorder, as the probability of the different length sequences start falling at a larger temperature ($T\approx 0.8$) than the HCP phase ($T\approx 0.3$). Even less robust to temperature disorder is the DHCP phase (not shown). In both regions, as temperature increases,  the non-stable phases creep in, being the DHCP (4H) phase, at any given temperature,  the one with higher probability of occurrence of the non-stable phases at that point. 

\section{Discussion\label{sec:discussion}}

IMLP raises some concerns regarding the physical grounds for writing an interaction Hamiltonian between fictitious entities. Furthermore, the simple use of a pairwise interaction model like the one described by (\ref{eq:isinghamiltonian}) is inconsistent as shown by the examples discussed in Section \ref{sec:math}. As discussed in RENL16, in order to lift such inconsistencies \cite{shaw90}, following \cite{cheng88}, have reported a more general Hamiltonian over spins as coded by the H\"agg code. This Hamiltonian considers, additionally to the sum of pairwise terms, the sum of terms made by monoids of an even number of spins. The introduction of such terms is rather artificial and cumbersome. Shaw and Heine go into a lengthy explanation for SiC, as why should additional terms be considered on the basis of geometrical arguments. Their conclusion is that in the IMLP the intermediate layers are important. As a consequence, the interaction parameters $J_n$ in front of the pairwise terms have no simple interpretation in terms of interaction between two layers, but have to account for numerous effects as those regarding intermediate layers. Yet, the introduction of non-pairwise terms is not made on physical grounds. The bottom line is that it seems that the introduction of such terms is made with the sole purpose of saving the H\"agg coded Ising Hamiltonian. It also follows that as longer range interactions are considered, the non-pairwise terms gets bigger and  harder to track mathematically.

This can be compared with the AKH, where a straight forward explanation, in physical terms, can be derived for the interaction parameters. In any case, involved arguments regarding cumbersome geometrical consideration of pair of layers is not needed. Even if it could be justified the need to take into account non-pairwise terms, AKH could accommodate such terms with little effort. 

It seems that, in absence of any other reason, just by an Occam's razor argument, AKH should be preferred to IMLP. But there is also the additional fact that the phase diagram obtained by each model are not equivalent: no invertible mapping can be established between both. At the end, one may ask if there is some experimental feature that can be explained by one of the models, and not by the other.

Both models predict the same stable phases at $T=0$, namely FCC, HCP and DHCP, difference must then be sought at the phase boundaries. In the FCC-DHCP boundary, all phases predicted by the IMLP are also predicted by the AKH, but not vice-versa. The AKH model predicts the occurrence, with probability larger than $5\%$, of the polytypes with Zhdanov symbol $1122$, $1232$, $17$, $1133$, $1125$, $1323$, $1243$, $1334$, $1235$ (Compare Table \ref{tbl:fccdhcp} in this work with Table I in RENL16). The same can be seen for the HCP-DHCP boundary. In this case, the AKH model predicts additionally the polytypes $1114$, $1232$, $1224$, $112123$, $11111212$ and $1212221$. In both boundaries, all the additional polytypes predicted by the AKH model are not close loop sequence in the $\epsilon$-machine FSM. The open loop polytypes have been associated with frozen sequence at the frontier between two sequence belonging to stable polytypes. When two polytipic sequences meet, both sequences can be out of phase at the boundary and therefore, one can not grow at the expense of the other. Arrested sequences are then formed. The AKH model predicts a larger number of arrested sequence than the IMLP at the FCC-DHCP and HCP-DHCP boundaries. It must be noted that all these arrested sequences are more complicated than the simple union of the two phases at a frontier. 

\cite{kiflawi76}, have reported in ZnS the experimental occurrence of the 8H polytypes $17$ and $1133$, which can be accommodate within the AKH model, but not the IMLP. Additionally Kiflawi et al reported the $(31)_3$ polytype which is not found in any of the two models. \cite{price84},  studying the application of the ANNNI model to polytype formation, cite experimental work by \cite{akaogi82} that reported in ZnS the occurrence of the $13$ structure. Yet, $13$ is not a valid polytype as it does not comply with the neutrality condition \cite{estevez05} and has to be considered a fragment of another polytype, or the true polytype $(13)_3$. In the first case, such fragment can happen in the $1213$ and the $111213$ polytypes reported by both models at the HCP-DHCP boundary or, in the $1133$, $1323$, $1334$ reported by the AKH model at the FCC-DHCP boundary. Other phases that Price and Yeomans report not being able to predict using the IMLP such as $233$ are also unaccounted by the AKH model, yet may be found as fragments in the $2233$ structure predicted by both models at the FCC-DHCP boundary. In an exhaustive list, up to the year 1972, of ZnS polytypes experimentally reported, Kailash enumerates nine polytypes up to length 12 \cite{kailash72}, of those  reported, only the occurrence of the $66$ polytype is not predicted by neither the IMLP or the AKH models.   

It is important to note, that additionally to predict all phases that the IMLP does, the AKH model predicts them with larger probability. Some phases predicted in the IMLP model have probabilities even below $1\%$ such as $1224$, $55$, $47$, $2324$, $2225$, $39$, $2343$, $2244$, $1213$, $111213$, $(12)_3$ among others,  some of them reported experimentally like the $55$ and the $2244$ \cite{kailash72}. It may be wondered if polytypes which such low probability could be observed with diffraction experiment technology in the early seventeens. For all these cases the AKH model predicts the occurrence of such polytypes with probabilities above $5\%$ and therefore more easily observable experimentally.

Regarding SiC, there are hundreds of polytypes reported\cite{verma66,ortiz12}, the three shortest ones, namely $3C$, $4H$, $6H$ can be predicted on the basis of the IMLP and the AKH model. The $8H$ polytype with Zhdanov symbol, $44$, and $10H$ with Zhdanov symbol $2332$, that have been reported experimentally, are also predicted by both models. All other reported polytypes have periodic length above $12$ and where not explored. 

\section{Conclusions}

Ising models based on the H\"agg coding of close packed sequence look like a ``natural'' choice and indeed has been used repetitively  in the past. The results of such models, in terms of the phase diagram and the allowance it makes of long period polytypes, seem to justify its use. Indeed, several experimental findings can well be accommodated within the model. Yet, an important issue has been raised by Ahmad and Khan on physical grounds that must be dealt with. It is indeed rather fictitious to write interaction terms between spins that represent layer displacements. What are the entities that interact? One way out has been suggested by adding monoid terms that comprises multiple ($>2$) layer interaction, but then this appears more like an attempt to fix a physically flawed approach. 

It has been shown that a Hamiltonian written in terms of direct interaction with the actual physical layers not only reproduces the occurrence of the same polytypes that the previous models, but also can justify the appearance of new structures, some of which have been experimentally found. In top of that, the interaction model is simpler than the Ising model over H\"agg spins, a simplicity that is even more evident as one scales to longer range interactions. Occam's rule should be applied. This Hamiltonian is not a result of this work, but has been introduced before by Ahmad and Khan.

What has been done in this contribution is to explore the implications of the AKH model for the phase diagram, the appearance of polytypes and the description of disorder, in terms of the machinery of computational mechanics. This has been done under the framework put forward by Crutchfield, Feldman, Varn and coworkers being already used in the context of polytype analysis. The reported FSM of the AKH, under different values of the parameters, allows to follow the balance between structure and disorder. The capability of the polytypic system to process and store information is given by the statistical complexity, while the emergence of structure is given by the excess entropy. Entropy density quantifies how much of irreducible randomness (structural disorder) there is in the system. Figure \ref{vsTFCC} and \ref{vsT} are important and exemplifies the type of analysis that can be performed once the $\epsilon$-machine describing the dynamics of the system has been deduced.

Also important has been the use of the FSM description to analyze the occurrence of different sequences in the stacking arrangement and its nature. Questions such as whether a  particular sequence can be considered an intrinsic polytype resulting of thermodynamic causes, or just an arrested frontier between stable phases are answered. This analysis would had been at least cumbersome by using the traditional tools of statistical physics. In this sense, computational mechanics develops tools that are complementary to other approaches.    

Crystallography is usually uncomfortable dealing with disorder, even when its partial. The usual approach is to disregard disorder all together, or to consider it as a perturbation of some underlying order. As already discussed by Varn et al in the reference given in this contribution, computational mechanics goes beyond the faulting model and fits disorder and order naturally within the same theoretical approach.

Finally, one could ask if Ising models should be disregarded completely. The authors believe that is to early to do so. Although some serious questions have been raised, both models still have to be systematically confronted with experimental evidence to a point that is still lacking. This paper only suggest that the AKH model can not be overlooked and must be taken seriously. We hope that our contribution motivates a renew interest in interaction models for studying polytypism.

\section{Acknowledgment}
 This work was partially financed by FAPEMIG under the project BPV-00047-13 and computational infrastructure support under project APQ-02256-12. EER which to thank AvH  for a fellowship renewal grant and the financial support under the PVE/CAPES grant 1149-14-8.  RLS wants to thank the support of CNPq through the projects 309647/2012-6 and 304649/2013-9.

\begin{table}
\caption{Polytypes of length $\leq 12$ with probability of occurrence above zero at the HCP-DHCP phase boundary, temperature is $T=0$($\beta=\infty$). The entries pointed by arrows corresponds to polytypes not generated by close loops in the corresponding $\epsilon$-machines FSM. L: polytype length, Prob: probability of occurrence normalized against all sequences of equal length, Ramsdell: Ramsdell notation, Zhdanov: Zhdanov symbol of the polytype, seq: the sequence coded in the ABC alphabet.}
\centering
\begin{tabular}{rllllrllll}
              L &  Prob.  & Ramsdell & Zhdanov   &  seq. &             L & Prob. & Ramsdell &   Zhdanov &  seq. \\
\hline\\
              2 &  0.450  &  2H   &  $11$        & AB         &               10 & 0.210 & 10H  & $11112112$    &  ABABABCBCB \\
              4 &  0.550  &  4H   &  $22$        & ABCB       &               10 & 0.173 & 10H  & $112222$      & ABABCBABCB \\
$\rightarrow$ 5 &  0.340  &  5R   &  $14$        & ABACB      & $\rightarrow$ 10 & 0.100 & 10H  & $112123$      & ABABCBCACB \\
              6 &  0.760  &  6H   &  $1122$      & ABABCB     &               10 & 0.090 & 10H  & $(122)_2$     & ABACABABCB\\
$\rightarrow$ 7 &  0.236  &  7R   &  $1213$      & ABACACB    &               11 & 0.139 & 11R  & $11121212$    &  ABABACACBCB \\
$\rightarrow$ 7 &  0.130  &  7R   &  $1114$      & ABABACB    & $\rightarrow$ 11 & 0.069 & 11R  & $11111212$    & ABABABACACB\\
              8 &  0.420  &  8H   &  $111122$    & ABABABCB   & $\rightarrow$ 11 & 0.062 & 11R  & $1212221$     & ABACACBCACB\\
              8 &  0.211  &  8H   &  $(112)_2$   & ABABCBCB   &               12 & 0.085 & 12H  & $11211222$    & ABABCBCBABCB \\
$\rightarrow$ 8 &  0.115  &  8H   &  $1232$      & ABACABCB   &               12 & 0.100 & 12H  & $1111111122$  & ABABABABABCB\\
$\rightarrow$ 9 &  0.180  &  9R   &  $111213$    & ABABACACB  &               12 & 0.100 & 12H  & $1111112112$  & ABABABABCBCB\\
$\rightarrow$ 9 &  0.100  &  9R   &  $1224$      & ABACABACB  &               12 & 0.088 & 12H  & $11212212$    & ABABCBCACBCB  \\
              9 &  0.096  &  9R   &  $(12)_3$    & ABACACBCB  &               12 & 0.087 & 12H  & $11122122$    & ABABACABABCB\\
             10 &  0.210  &  10H  &  $11111122$  & ABABABABCB &               12 & 0.085 & 12H  & $11112222$    &  ABABABCBABCB \\
\end{tabular}\label{tbl:hcpdhcp}
\end{table}

\begin{table}
\caption{Polytypes of length $\leq 12$ with probability of occurrence above zero at the FCC-DHCP phase boundary, temperature is $T=0$($\beta=\infty$). Conditions and notation follow Table \ref{tbl:hcpdhcp}.}
\centering
\begin{tabular}{rllll}
	               L & Prob. & Ramsdell & Zhdanov &  seq.\\
	\hline\\
	               3 & 0.720 & 3C  & $\infty$ &  ABC\\
	               4 & 0.550 & 4H  & $22$     &  ABCB\\
	 $\rightarrow$ 5 & 0.620 & 5R  & $14$     & ABACB\\
	               6 & 0.410 & 6H  & $33$     &  ABCACB\\
	 $\rightarrow$ 6 & 0.211 & 6H  & $1122$   &  ABABCB\\
	               7 & 0.58  & 7R  & $25$     & ABCBACB\\
	               8 & 0.211 & 8H  & $44$     & ABCABACB\\
	 $\rightarrow$ 8 & 0.146 & 8H  & $17$     & ABACBACB\\
	 $\rightarrow$ 8 & 0.115 & 8H  & $1232$   & ABACABCB\\
	 $\rightarrow$ 8 & 0.081 & 8H  & $1133$   & ABABCACB\\
	               9 & 0.310 & 9R  & $36$     & ABCACBACB \\
	 $\rightarrow$ 9 & 0.161 & 9R  & $1224$   & ABACABACB \\
	 $\rightarrow$ 9 & 0.100 & 9R  & $1125$   & ABABCBACB \\
	
\end{tabular}\;\;\; \;\;
\begin{tabular}{rllll}
	  L & Prob. & Ramsdell & Zhdanov &  seq. \\
	  \hline\\

	  $\rightarrow$  9 & 0.090 & 9R   & $1323$  & ABACBCACB \\
	                10 & 0.210 & 10R  & $28$    & ABCBACBACB \\
	                10 & 0.180 & 10H  & $2233$  & ABCBABCACB \\
	                10 & 0.106 & 10H  & $55$    & ABCABCBACB \\
	  $\rightarrow$ 10 & 0.100 & 10R  & $1243$  & ABACABCACB \\
	                11 & 0.146 & 11R  & $47$    & ABCABACBACB \\
	                11 & 0.125 & 11R  & $2324$  & ABCBACABACB \\
	                11 & 0.122 & 11R  & $2225$  & ABCBABCBACB \\
	  $\rightarrow$ 11 & 0.069 & 11R  & $1334$  & ABACBCABACB \\
	  $\rightarrow$ 11 & 0.062 & 11R  & $1235$  & ABACABCBACB \\
	                12 & 0.101 & 12R  & $39$    & ABCACBACBACB \\
	                12 & 0.088 & 12R  & $2343$  & ABCBACABCACB \\
	                12 & 0.085 & 12H  & $2244$  & ABCBABCABACB \\
\end{tabular}\label{tbl:fccdhcp}
\end{table}


\end{document}